\documentclass[fleqn,10pt]{wlscirep}
\title{Crystal Structure of 200 K-Superconducting Phase of Sulfur Hydride System}

\author[1,*,+]{Mari Einaga}
\author[1]{Masafumi Sakata}
\author[1]{Takahiro Ishikawa}
\author[1,+]{Katsuya Shimizu}
\author[2,+]{Mikhail I. Eremets}
\author[2]{Alexander P. Drozdov}
\author[2]{Ivan A. Troyan}
\author[3]{Naohisa Hirao}
\author[3]{Yasuo Ohishi}

\affil[1]{KYOKGEN, Graduate School of Engineering Science, Osaka university, Machikaneyamacho 1-3, Toyonaka, Osaka, 560-8531, Japan}
\affil[2]{Max-Planck Institut f\"{u}r Chemie, Hahn-Meitner-Weg 1, 55128 Mainz, Germany}
\affil[3]{JASRI/SPring-8, 1-1-1, Sayo-cho, Sayo-gun, Hyogo 679-5198. Japan}

\affil[*]{einaga@hpr.stec.es.osaka-u.ac.jp}

\affil[+]{these authors contributed equally to this work}

\newcommand{\Tc}{\ensuremath{T_\mathrm{c}}}

\begin{abstract}
\large
Superconductivity with the critical temperature \Tc  \ above 200 K has been recently discovered by compression of H$_2$S (or D$_2$S) at pressure above 90 GPa.
It was proposed that these materials decompose under pressure to elemental sulfur and hydride with higher content of hydrogen\cite{Drozdov2014,Drozdov2015} which is responsible for the high temperature superconductivity.
Here we report the crystal structure of the superconducting compressed H$_2$S and D$_2$S by synchrotron x-ray diffraction (XRD) measurements combined with electrical resistance measurements at room and low temperatures. 
We found that the superconducting (SC) phase is mostly in good agreement with theoretically predicted hexagonal and body-centred cubic (bcc) structure \cite{Duan2014}, and coexists with elemental sulfur, which claims that the formation of 3H$_2$S $\rightarrow$ 2H$_3$S + S is occur by the compression \cite{Duan2014,Duan2015,Bernstein2015,Errea2015}.

\end{abstract}

\begin{document}

\flushbottom
\maketitle
%
%

\Large

Recently very high \Tc  \ of 200 K has been discovered in hydrogen sulfide. The SC was proved by the sharp drop of resistance to zero, a strong isotope effect in study of D$_2$S, the shift of SC transition with magnetic field and finally in the studies of magnetic susceptibility and magnetization.
The structure of the SC phase was not experimentally measured but it was theoretically proposed that most likely the staring H$_2$S decomposes under pressure (with assistance of temperature) to pure sulfur and some sulfur hydride with higher content of hydrogen (such as SH$_4$ or similar to that). 
At the same time with a theoretical work\cite{Duan2014} appeared which considered a different starting material (H$_2$S)$_2$H$_2$ (stoihiometry H$_3$S) and found  $R$3$m$ and $Im$-3$m$ structures under pressure above 111 GPa and 180 GPa, respectively.
These structures were further carefully studied theoretically by different methods and \Tc\ $\sim$ 200 K was consistently calculated for the $Im$-3$m$ structure. 
The calculated \Tc \  as well as its pressure dependence are close to the experimental data\cite{Drozdov2014,Drozdov2015}. 
This agreement suggests that the high \Tc \ observed in the experiments relate to the H$_3$S with the $Im$-3$m$ structure. 
Later calculations supported this idea: H$_2$S is unstable at high pressures and decomposes to sulfur and higher hydrides, most likely to H$_3$S\cite{Bernstein2015,Errea2015}. 
The goal of the present work is to determine the structure of the superconducting hydrogen sulfide and verify, if it really corresponds to the theoretically predicted structure.

Figure 1a shows a typical x-ray diffraction image of sulfur hydride at room temperature and the integrated intensity which excludes the background.
It is considered that the reflections from the sample have the anisotropy in intensity is caused by the inhomogeneous of them. 
The main pattern well corresponds to the theoretically predicted bcc and hexagonal phase of H$_3$S \cite{Duan2014}.
Then the pattern can be explained by the mixture of H$_3$S and $\beta$-polonium (Po) structure of elemental sulfur \cite{Luo1993} except the peaks marked with triangle ($\triangledown$), star ($\ast$) and open circle ($\circ$).
We also carried out XRD measurements in sulfur deuteride under high pressure as shown in Fig. 1b. 
Structures of sulfur deuteride are the same - most of the reflections can be explained by the mixture phase of bcc and $\beta$-Po structure phases. 
But no peak was observed at around 13.3 degree in any sulfur deuteride which emerges in sulfur hydride ($\ast$).

We have performed that XRD and electrical resistance measurements simultaneously in sulfur deuteride at 173 GPa. Top of the fig. 1b shows the temperature dependence of the XRD patterns.
The XRD patterns kept at all measured temperature region and the resistance measurement shows the drop indicates superconducting transition below 150 K (Fig. 3b). 
Therefore, the structure in the normal and SC states are the same - no visible structural transition at SC transition.

As shown in Fig. 1c, the obtained pattern and the calculated that which is result of Rietveld refinement using RIETAN-FP \cite{Izumi2007} with the predicted structure in ref. \cite{Duan2014} show the good agreement as follows: $R_\mathrm{wp}$ = 0.40\%, $S$ = 1.92 in bcc lattice and $R_\mathrm{wp}$ = 0.41\%, $S$ = 1.94 in hexagonal lattice. It is because that both of them are of same periodicity in sulfur atoms of H$_3$S, which have bcc lattice. And these lattice parameters are $a_\mathrm{bcc}$ = 3.089 (1) \AA \ in bcc lattice and $a_\mathrm{hex}$ = 4.379 (10) \AA, $c_\mathrm{hex}$ = 2.663 (9) \AA \ in hexagonal lattice. That of $\beta$-Po sulfur are $a$ = 3.696 (1) \AA, $c$ = 2.708 (1) \AA.
In sulfur deuterides at 173 GPa at room temperature, lattice parameters are $a_\mathrm{bcc}$ = 3.0468 (2) \AA, $a$ = 3.351 (3) \AA \   and $c$ = 2.659 (3) \AA \ with $R_\mathrm{wp}$ = 0.28\%, $S$ = 0.55.

The XRD patterns obtained on decreasing process from 150 GPa to 92 GPa in sulfur hydride are shown in Fig. 2a.
No structural change was observed in main peaks. 
The intensity of the reflections come from $\beta$-Po structural elemental sulfur decline with decreasing pressure, and then disappear at 103 GPa. 
Meanwhile, a peak around 12.2 degree ($\circ$) which perhaps is incommensurate phase IV of elemental sulfur enhanced.
The undefined peak ($\bullet$) around 9.5 degree disappear at 111 GPa. 
It is considered that other peaks around 10.5 degree ($\triangledown$) and 13.3 degree ($\ast$) are from possible minor phase because these peaks shifted with the main peaks by pressure.
The XRD patterns of sulfur deuteride obtained on increase and decrease from 173 GPa to 190 GPa then 180 GPa are shown in Fig. 2b. The pattern doesn’t change by pressure except the shift of peak positions with pressure changes.

Pressure dependence of the atomic volume, $V_{atm}$, of sulfur hydride and sulfur deuteride calculated with predicted \cite{Duan2014} bcc-structures are shown in Fig. 2c. 
The compression curve (solid line) was fit using three experimental points, 150 GPa (sulfur hydride), 173 GPa and 190 GPa (sulfur deuteride).
The other points obtained by pressure-decreasing run may show it incompressible.
The solid line is drawn by the first-order Birch equation of state \cite{Birch1978} with the bulk modulus $B_{0}$ = 486 (35) GPa, and its pressure derivative $B_{0}$' = 4 (fixed).
The value of experimentally observed $V_{atm}$ is slightly big and the compressibility shows good agreements with the theoretically predicted one \cite{Duan2014}.

Pressure dependence of the normalized atomic volume $V/V_{0}$ of $\beta$-Po structural elemental sulfur is shown in Fig. 2d. Closed circles indicate the experimental results of Luo, which are fit by the fist-order Birch EOS (solid line) \cite{Luo1993}.
The deviation of the present work from the result of Luo can be considered that the lattice is expanded due to hydrogenation of $\beta$-Po sulfur which is generated from the dissociation of H$_2$S.
As shown in broken line, the calculated result of pressure dependence of $V/V_{0}$ in SH$_{1/27}$ which has $\beta$-Po structure is able to explain the experimental result.

The temperature dependencies of the electrical resistance of sulfur hydride and deuteride measured simultaneously with XRD patterns (Fig. 1a and 1b) are shown in Fig. 3a and 3b.
The summary of the values of \Tc \ in present and previous works are plotted in Fig. 3c. 
The present data agree with the previous data \cite{Drozdov2015} above 133 GPa.
\Tc \ suddenly decreased from 123 GPa and settled down. 
It seems that a phase transition from high-\Tc \ phase to low-\Tc \ occurs. 
In fact, the resistance at room temperature also increases suddenly at the pressure (Fig. 3a), but no evidence of this structural phase transition is detected in XRD measurements.

As Drozdov and Fujihisa have reported in ref. \cite{Drozdov2015,Fujihisa2004}, molecular hydrogen (H$_2$) was not observed in the sample chamber by both Raman scattering and XRD measurements. 
The crystal structures of  main phase of sulfur hydride and sulfur deuteride have the bcc-lattice ordering of sulfur atom. 
Thus, it is conceivable that H$_2$S/D$_2$S molecular are decomposed into H$_3$S/D$_3$S molecular which has bcc structure and elemental sulfur by compression as predicted by theoretical work. 
We claim that the bcc structural H$_3$S/D$_3$S corresponds to high-\Tc \ phase over 200 K. 
Note however that the dissociated H$_2$S/D$_2$S may supply a slight amount of hydrogen atom to sulfur, and the bcc-structural H$_3$S/D$_3$S lose the hydrogen atoms like atomic vacancy. 
Taking into account the present theoretical work, it is expected that the chemical reaction occurs as follows: 3H$_2$S $\rightarrow$ 2H$_{3-\delta}$S + SH$_{2\delta}$, ($2\delta \sim 1/27$). 
The investigation of the undefined minor phases is now going on.






 



\section*{Methods}

The sample preparations in DACs are almost same setup with those of ref. \cite{Drozdov2015} with diamond mount with wider diffraction angle.
Angle-dispersive powder x-ray diffraction measurements were carried out in beamline 
BL10XU of the SPring-8. The sample in the DACs was irradiated using synchrotron 
radiation beams monochromatized to energy of 30.0 keV ($\lambda$ $\sim$ 0.412 $\--$ 0.414 
\AA). Simultaneous measurements of XRD and electrical resistance were carried out in 
D$_2$S at 173 GPa under low temperature region, 13 - 300 K. DAC set in a cryostat. Each 
diffraction pattern was recorded using an imaging plate with an exposure time between 
120 and 300 seconds. The electrical resistance was measured by a commercial AC-resistance bridge (Linear Research Inc., LR-700) with four probe method.

We determined the amount of hydrogen impurity in $\beta$-Po sulfur by comparing pressure-volume curve between experiments and first-principles calculations. 
The Quantum ESPRESSO code \cite{giannozzi2009quantum} was utilized for the calculations based on the density functional theory, in which the Perdew-Burke-Ernzerhof generalized gradient approximation \cite{perdew1996generalized} was used for the exchange-correlation functional and the Vanderbilt ultrasoft pseudopotential \cite{Vanderbilt:1990kq} was employed. 
We calculated the pressure-volume curve of $\beta$-Po sulfur with hydrogen impurity by setting a hydrogen atom in the supercell of sulfur and performing the structural optimization.
The position of the hydrogen atom, the size of the supercell, and the number of k-points are as follows: (i) $(1/4, 1/4, 1/4), 2 \times 2 \times 2$ primitive cells, and $12 \times 12 \times 12$, and (ii) $(1/6, 1/6, 1/6), 3 \times 3 \times 3$ primitive cells, and $8 \times 8 \times 8$. The energy cut-off of the plane wave basis was set at 80 Ry.



\section*{Acknowledgements}

This work was concluded under proposal No. 2015A0112 of the SPring-8. This research was supported by Japan Society for the Promotion of Science Grant-in-Aid for Specially Promoted Research, No. 26000006, JSPS KAKENHI, Grant-in-Aid for Young Scientists (B), No.15K17707 and the European Research Council 2010-Advanced Grant 267777.

\section*{Author contributions}
M.E. performed the whole XRD measurement and the data interpretation and writing the manuscript. M.S. performed the cryogenic operations and XRD date collections. K.S. performed the in situ electrical resistance measurements in XRD measurements and writing the manuscript. T.I. performed the support calculations for the data interpretation. M.I.E. designed the study and participated in XRD experiments. A.P.D. and I.A.T. prepared the sample in DAC for whole experiments. N.H. and Y.O. performed the optimization of synchrotron XRD and cryogenic operations. M.E., K.S. and M.I.E. contributed equally to this paper.

\section*{Competing financial interests}The authors declare that they have no competing financial interests.

\section*{Correspondence} Correspondence and requests for materials should be addressed to M. E. (email: einaga @hpr.stec.es.osaka-u.ac.jp).

\begin{figure}[ht]
\centering
\includegraphics[width=\linewidth]{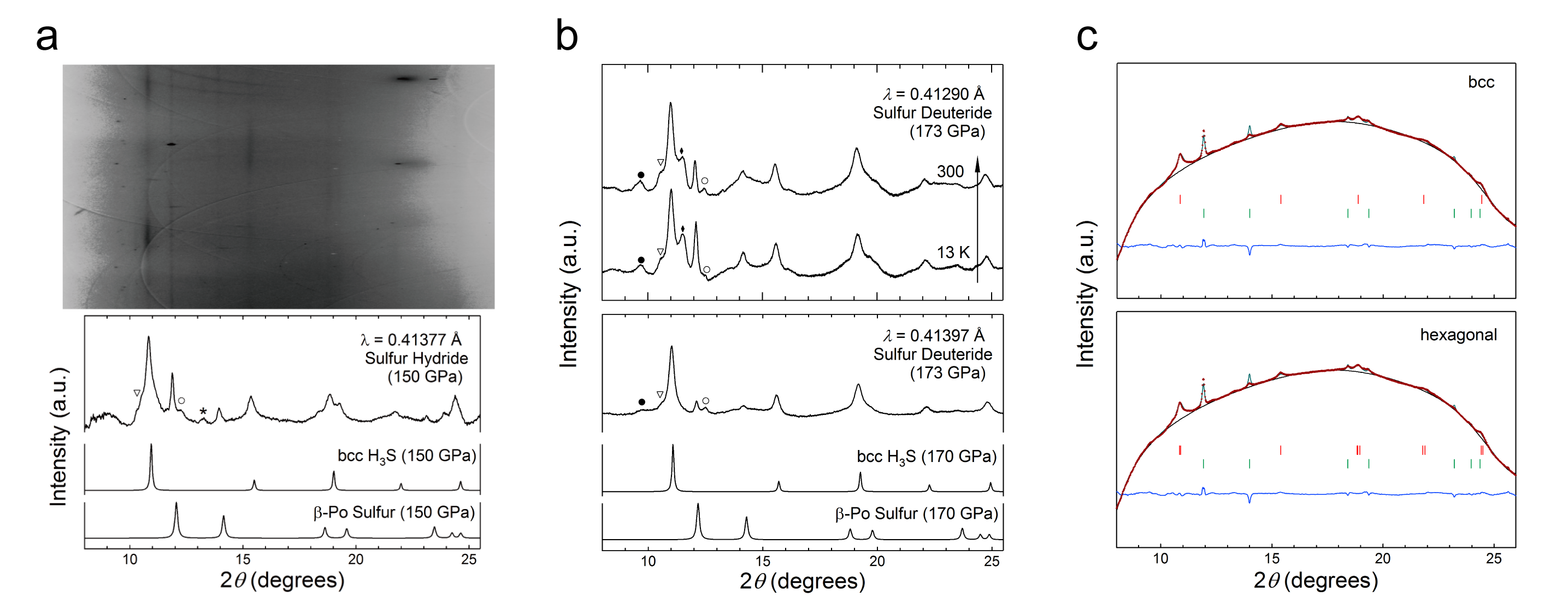}
\label{fig:1}
\caption{$\textbf{XRD measurement in sulfur hydride and sulfur deuteride systems.}$ \textbf{a}, (top) Unrolled powder diffraction image of sulfur hydride at 150 GPa at room temperature recorded on imaging plate. These arcuate lines and dots come from the Kossel line from diamond anvils. (bottom) X-ray diffraction pattern which is obtained by integration of top, which excludes background. The marks of open triangle and star indicate the undefined minor phase. Open circle indicates the reflection from other high pressure phase. Calculated patterns of bcc H$_{3}$S and $\beta$-Po elemental sulfur at 150 GPa from ref. \cite{Duan2015,Luo1993} lie below.
\textbf{b}, (top) Temperature dependence of XRD patterns of sulfur deuteride at 173 GPa. Dot and open triangle are reflection comes from undefined minor phase. Diamond are from gasket. Open circle indicates the reflections from other high pressure phases. (bottom) X-ray diffraction pattern at room temperature and 173 GPa which excludes background.
\textbf{c}, Rietveld fit of diffraction pattern of Fig. 1a with bcc/hexagonal structural H$_{3}$S and $\beta$-Po elemental sulfur at 150 GPa and room temperature. Red Dots, green and black lines represent the observed, calculated intensity and back ground, respectively. Blue line at the bottom indicates the residual error. Top and bottom ticks are peak position of the reflection from H$_{3}$S and elemental sulfur, respectively.}
\end{figure}

\begin{figure}[ht]
\centering
\includegraphics[width=\linewidth]{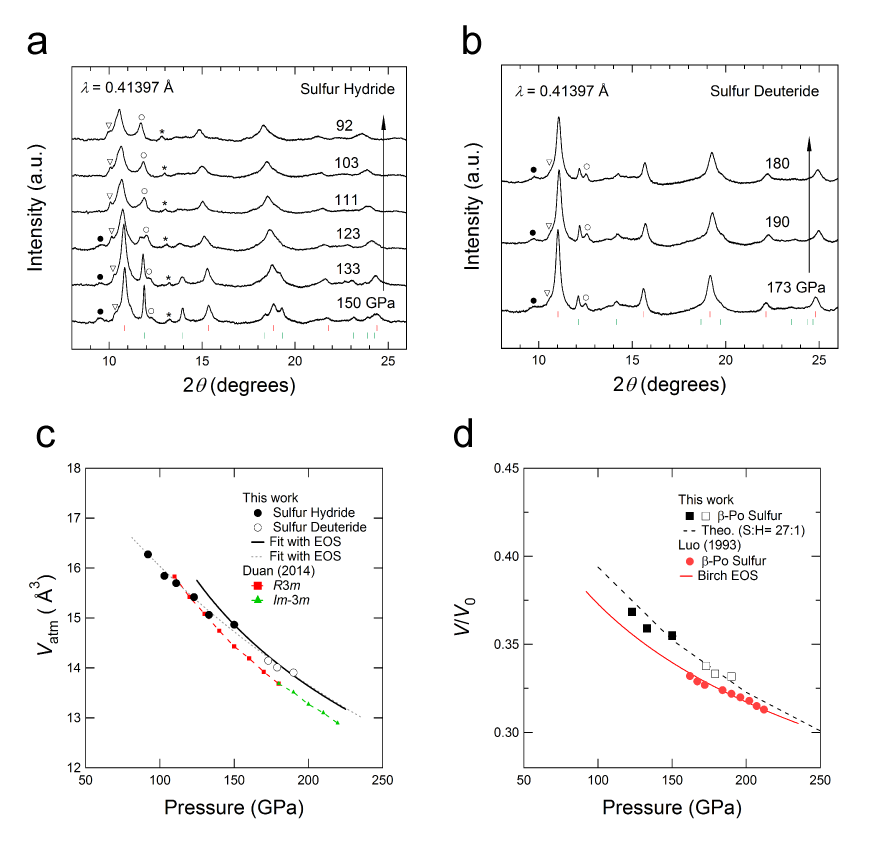}
\caption{$\textbf{Pressure dependence of XRD patterns in sulfur hydride and sulfur deuteride systems.}$ \textbf{a}, Pressure dependence in decreasing pressure of XRD patterns in pressurized sulfur hydride at room temperature. Upper (red) and lower (green) ticks indicates the peak position of predicted bcc structure and $\beta$-Po elemental sulfur, respectively. The peaks marked with dot and open triangle, and star are undefined minor phase. Open circles indicate the reflection from other high pressure phase.
\textbf{b}, Pressure dependence in decreasing pressure of XRD patterns in pressurized sulfur deuteride at room temperature. Upper (red) and lower (green) ticks indicates the peak position of predicted bcc structure and $\beta$-Po structural elemental sulfur, respectively. The peaks marked with star are undefined peak. Dot and open circle indicate the reflection from other high pressure phase.
\textbf{c}, Pressure dependence of the atomic volume of sulfur hydride and deuteride which are marked with solid and open symbol, respectively. Experimentally obtained data are fit with first-order Birch EOS indicated with the black solid line (increasing process) and dotted line (all). The theoretical predicted volumes of hexagonal ($R$3$m$) and bcc ($Im$-3$m$) structures are shown in solid square ($\blacksquare$) and triangle ($\blacktriangle$) with broken line, respectively \cite{Duan2014}.
\textbf{d}, Pressure dependence of the normalized volume $V$/$V_0$ in $\beta$-Po structural elemental sulfur ($\blacksquare$, $\Box$). Broken line indicates the simulated $V$/$V_0$ which contains 1/27 hydrogen atoms per a sulfur atom in the present work. Close circle indicates the experimentally obtained volume and solid line is the fitting curve with first-order Birch EOS with $B_0$ = 30.6 GPa, $B_{0}$' = 6, $V_0$ = 25.64 \AA$^3$ in ref. \cite{Luo1993}. }
\label{fig:2}
\end{figure}

\begin{figure}[ht]
\centering
\includegraphics[width=\linewidth]{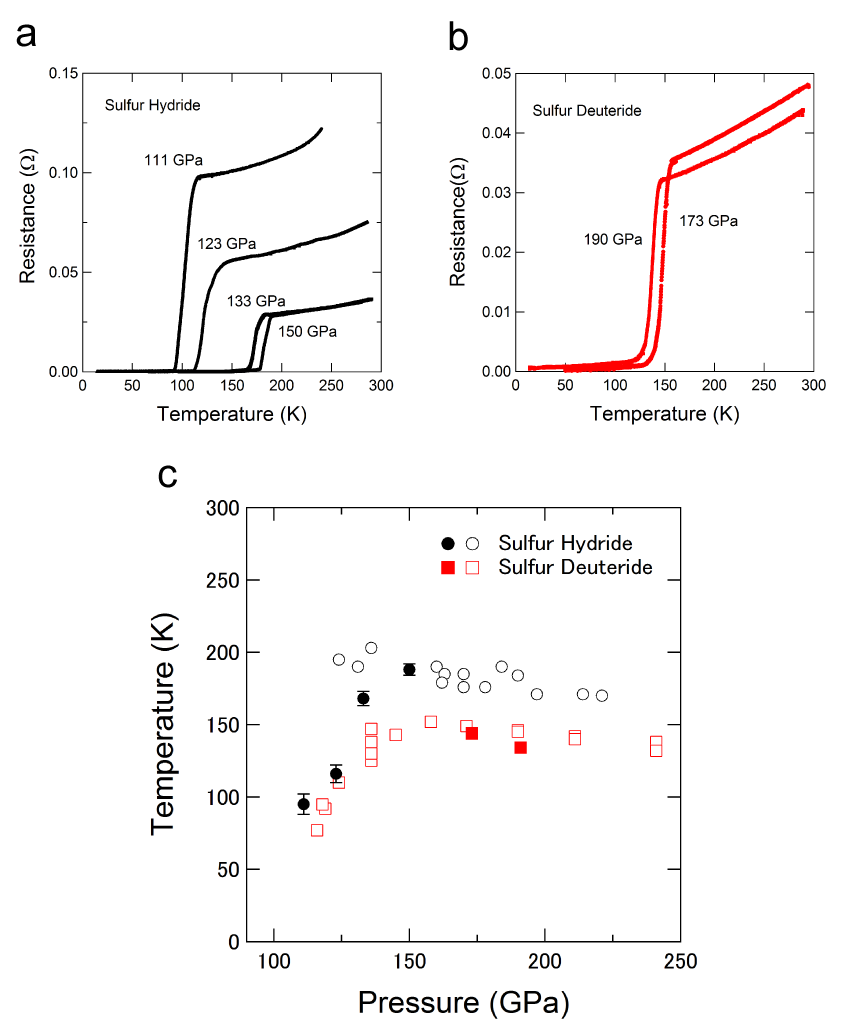}
\caption{$\textbf{Pressure dependence of $T_\mathrm{c}$ in sulfur hydride and sulfur deuteride.}$ \textbf{a}, Temperature dependence of resistance in sulfur hydride on heating process. \textbf{b}, and sulfur deuteride. \textbf{c}, Pressure dependence of \Tc \ of sulfur hydride and sulfur deuteride which are marked with circle and square symbol, respectively. Open circles, square and triangles are obtained in increasing pressure process \cite{Drozdov2015}. Closed circles and squares are the data in increasing and decreasing pressure that of this work, respectively.}
\end{figure}

\end{document}